\documentclass[twocolumn,showpacs,preprintnumbers,amsmath,amssymb]{revtex4}
\usepackage{epsfig}

\usepackage[table]{xcolor}
\usepackage{graphicx}% Include figure files
\usepackage{dcolumn}% Align table columns on decimal point
\usepackage{bm}% bold math
\usepackage{esvect}

\usepackage{hyperref}
\hypersetup{colorlinks=true,linkcolor=magenta,anchorcolor=green,citecolor=cyan,filecolor=black,menucolor=black,urlcolor=brown}
\usepackage{slashed}
\newcommand{\be}{\begin{equation}}
\newcommand{\ee}{\end{equation}}
\newcommand{\ba}{\begin{eqnarray}}
\newcommand{\ea}{\end{eqnarray}}

\begin{document}

\title{Fate of scalar dark matter solitons around supermassive galactic black holes}

\author{Philippe Brax}
\affiliation{Institut de Physique Th\'eorique, Universit\'e  Paris-Saclay, CEA, CNRS, F-91191 Gif-sur-Yvette Cedex, France}
\author{Jose A. R. Cembranos}
\affiliation{Departamento de  F\'{\i}sica Te\'orica and IPARCOS,\\
Universidad Complutense de Madrid, E-28040 Madrid, Spain}
\author{Patrick Valageas}
\affiliation{Institut de Physique Th\'eorique, Universit\'e  Paris-Saclay, CEA, CNRS, F-91191 Gif-sur-Yvette Cedex, France}

\begin{abstract}
In scalar-field dark matter scenarios, a scalar-field soliton could form at the center
of galactic halos, around the supermassive black holes that sit at the center of galaxies.
Focusing on the large scalar-mass limit, where the soliton is formed by the balance
between self-gravity and a repulsive self-interaction, we study the infall of the scalar field
onto the central Schwarzschild black hole. We derive the scalar-field profile, from
the Schwarzschild radius to the large radii dominated by the scalar cloud.
We show that the steady state solution selects the maximum allowed flux, with a critical
profile that is similar to the transonic solution obtained for the hydrodynamic case.
This finite flux, which scales as the inverse of the self-interaction coupling, is small enough
to allow the dark matter soliton to survive for many Hubble times.

\end{abstract}

\date{\today}% It is always \today, today,

\maketitle

%\tableofcontents

\section{Introduction}
\label{sec:introduction}

In the last few years, there has been an increasing number of theoretical and experimental analyses investigating  the fundamental nature of Dark Matter (DM). A possible candidate which has attracted  a lot of
attention  is related to ultralight coherent fields. This idea is not new and
dates back to the pioneering studies of the QCD axion as a DM candidate
\cite{Peccei:1977hh, Wilczek:1977pj, Weinberg:1977ma}. More generally, Axion-Like-Particles (ALPs) are well motivated
by different theories \cite{Marsh:2015xka, Hui:2016ltb} with a broad range of masses and couplings
\cite{Sarkar:2015dib,  Kobayashi:2017jcf, Abel:2017rtm,Banik:2017ygz, Hirano:2017bnu, Conlon:2017ofb,
Brito:2017wnc, Brito:2017zvb,Sarkar:2017vls, Diacoumis:2017hff}.

The cosmological interest of these DM candidates is associated with the fact that their de Broglie wavelength is
of the order of astrophysical scales \cite{Hu:2000ke}. This type of coherent DM is constituted of fast
oscillating massive scalars  \cite{Turner:1983he, Johnson:2008se} or other higher-spin bosonic fields
\cite{Cembranos:2012kk,Cembranos:2012ng,Cembranos:2013cba,Alvarez-Luna:2018jsb}. For large distances,
both at the background and perturbation levels, these coherent fields behave as Cold DM (CDM) does
\cite{Johnson:2008se, Hwang:2009js, Park:2012ru, Hlozek:2014lca, Cembranos:2015oya,Cembranos:2016ugq}.
However, at shorter scales, the matter power spectrum is highly suppressed \cite{Hlozek:2014lca} and the
the formation of cusps is replaced by DM cores \cite{Schive:2014dra, Broadhurst:2018fei}.
These distinctive features of the structure formation associated with light coherent bosonic fields
have attracted a lot of  attention  due to longstanding tensions between standard CDM predictions
and different observations on galactic and sub-galactic scales \cite{Ostriker:2003qj,Cembranos:2005us,Weinberg:2013aya,Pontzen:2014lma,BoylanKolchin:2011de,Moore:1999nt,deBlok:2009sp,Cembranos:2018ulm}.

In particular, the cored density profiles that this type of DM develops
\cite{Arbey:2001qi,Lesgourgues:2002hk,Chavanis:2011zi,Chavanis:2011zm,Schive:2014dra,Schive:2014hza,Marsh:2015wka,Calabrese:2016hmp,Chen:2016unw,Schwabe:2016rze,Veltmaat:2016rxo,Hui:2016ltb,Gonzales-Morales:2016mkl,Robles:2012uy,Bernal:2017oih,Mocz:2017wlg,Mukaida:2016hwd,Vicens:2018kdk,Bar:2018acw,Eby:2018ufi,Bar-Or:2018pxz,Marsh:2018zyw,Chavanis:2018pkx,Emami:2018rxq,Levkov:2018kau,Broadhurst:2019fsl,Hayashi:2019ynr,Bar:2019bqz}, are referred to as solitons. They correspond to stationary or steady solutions of the
classical field equations of the bosonic field. In this context, it appears to be  necessary to take into account that most galaxies
host a supermassive Black Hole (BH) in their central region \cite{Kormendy:1995er,Ferrarese:2004qr,Narayan:2005ie}.
The dynamics and phenomenology of ultralight scalar fields within the geometry sourced by a BH have been studied in
 \cite{Ferreira:2017pth,Boskovic:2018rub,Cardoso:2018tly,Davoudiasl:2019nlo,Hui:2016ltb,Bar:2018acw,Bar-Or:2018pxz,Desjacques:2019zhf,Bar:2019pnz,Hui:2019aqm}.

In this work, we analyze the impact of anharmonic self-interactions on scalar DM solitons and their fate in the presence of a central BH.
We focus on the quartic case where oscillations of the scalar field are given in terms of elliptic functions that reduce to trigonometric functions in the absence of interactions.  These anharmonic corrections introduce large deviations with respect to the
standard CDM scenario. They source additional effective pressure (positive for the repulsive case
\cite{Goodman:2000tg,Li:2013nal, Suarez:2016eez,Suarez:2015fga, Suarez:2017mav, Chavanis:2018pkx}
and negative for the attractive one \cite{Cedeno:2017sou,Desjacques:2017fmf}), which may alleviate the
 small scale problems of CDM \cite{Fan:2016rda} and lead to the existence of  vortices in galaxies \cite{RindlerDaller:2011kx}.
These modifications can be also used to constrain the parameter space of ultralight coherent DM.
In fact, effects on the CMB anisotropies \cite{Cembranos:2018ulm}, large-scale structures \cite{Cembranos:2018ulm}
and gravitational waves \cite{Dev:2016hxv,Li:2016mmc} have been already considered for this purpose.

Here we focus on the scalar field profile and behavior around a central BH. We find that outside the Schwarzschild radius and close enough to the Black Hole the scalar dynamics are described  by a stationary solution with non-vanishing flux. This corresponds to the in-fall of dark matter into the central BH. Far away from the center, the dynamics reproduce the static soliton behavior, with a solution whose density is nearly constant in the core before falling off rapidly towards zero \cite{Brax:2019fzb}. This selects a unique solution with constant flux and nearly vanishing velocity far away from the BH which is similar to the transonic solution obtained for the hydrodynamic case.
We find typically that the lifetime of the soliton, despite the falling of matter into the BH, is larger than the age of the Universe. Moreover the constraints on the density profile of dark matter inferred from the stellar dynamics in the vicinity of the central BH \cite{Yu:2016nzn,Akiyama:2019eap} are easily met.

This manuscript is arranged as follows.
In section~\ref{sec:dark-matter}, we describe the main equations of a generic model of scalar DM
within a Schwarzschild geometry, both in Isotropic coordinates~\ref{sec:isotropic}
and Eddington coordinates ~\ref{sec:Eddington}.
In section~\ref{sec:free-scalar-field}, we analyze the main features of the
scalar DM solitons for the harmonic case.
In section~\ref{sec:quartic}, we extend this analysis to the self-interacting case determined by
a quartic term.
In section~\ref{sec:lifetime}, we derive the long lifetime associated with the scalar-field soliton
found in the previous section.
Finally, the main conclusions are summarized in section~\ref{sec:conclusion}.

\section{Dark matter scalar field}
\label{sec:dark-matter}

The scalar-field action is
\be
S_\phi = \int d^4x \sqrt{-g} \left[ - \frac{1}{2} g^{\mu\nu} \partial_\mu\phi \partial_\nu\phi
- V(\phi) \right] .
\ee
We also write the scalar-field potential as
\be
V(\phi) = \frac{m^2}{2} \phi^2 + V_{\rm I}(\phi) ,
\label{eq:V-def}
\ee
where $V_{\rm I}$ is the self-interaction potential.
In this work we focus on the quartic self-interaction potential,
\be
V_{\rm I}(\phi) = \frac{\lambda_4}{4} \phi^4 .
\label{eq:V-quartic}
\ee
Such scalar fields can play the role of DM and build scalar solitons,
i.e. static profiles with a finite core, at the center of galactic halos.
These solitons can be the result of the balance between the self-gravity of the
scalar cloud and a ``quantum pressure'' (due to the fact that the underlying
equations of motion are the Klein-Gordon equation, or the Schr\"odinger equation
in the nonrelativistic limit, rather than the hydrodynamical Euler equation),
or to a repulsive self-interaction, associated with $\lambda_4 > 0$.
In this paper, following our previous work \cite{Brax:2019fzb}, we focus
on the large scalar mass limit,
\be
m \gg 10^{-21} \, {\rm eV} ,
\label{eq:m-cosmo}
\ee
which ensures that the quantum pressure is negligible from cosmological to galactic
scales.
Then, the galactic solitons are due to the balance between gravity and the repulsive
self-interaction.
In the large scalar mass limit, the analysis simplifies and we can derive in the next
sections explicit expressions for the scalar field profile and its inflow onto the supermassive
BH. Around Schwarzschild BH, we shall see below that the large-mass
limit becomes defined by the lower bound (\ref{eq:m-BH}), which is somewhat
larger than (\ref{eq:m-cosmo}).

In this work, we focus on spherically symmetric systems, as we consider
a spherical scalar cloud around a supermassive Schwarzschild BH.

\subsection{Isotropic coordinates}
\label{sec:isotropic}

Close to the BH, the contribution from the scalar field is negligible
and the metric is the standard Schwarzschild metric
\cite{Poisson:2009pwt,Blau-2017}
\be
ds^2 = - \left(1-\frac{r_s}{\tilde r} \right) dt^2 + \left(1-\frac{r_s}{\tilde r}\right)^{-1} d\tilde r^2
+ \tilde r^2 d\vec\Omega^2 ,
\label{eq:ds2-Schwarzschild}
\ee
where $\tilde r$ is the Schwarzschild radial coordinate
and $r_s=2{\cal G}M/c^2$ is the Schwarzschild radius of the BH of mass $M$.
To simplify the matching with the Newtonian gauge at large scales, in the following
we work with the isotropic radial coordinate $r$ \cite{Blau-2017}, which is related to the
Schwarzschild radial coordinate by
\be
\tilde{r} > r_s, \;\; r > \frac{r_s}{4} : \;\;\;
\tilde r= r \left(1+\frac{r_s}{4r}\right)^2 ,
\ee
and the Schwarzschild metric becomes
\be
ds^2 = - f(r) dt^2 + h(r) ( dr^2 + r^2  d\vec\Omega^2 ) ,
\label{eq:ds2-Schwarzschild-isotropic}
\ee
with
\ba
\frac{r_s}{4} < r \ll r_{\rm sg} : && f(r) = \left( \frac{1-r_s/(4r)}{1+r_s/(4r)}
\right)^2 , \nonumber \\
&& h(r) = (1+r_s/(4r))^4 .
\label{eq:f-h-def}
\ea
The metric (\ref{eq:ds2-Schwarzschild-isotropic})-(\ref{eq:f-h-def}) applies at radii
$r \ll r_{\rm sg}$,
where $r_{\rm sg}$ is the transition radius where the self-gravitational contribution
to the metric potentials from the DM, that is, the scalar cloud, becomes important.
This corresponds to the radius where metric fluctuations have decreased down to
$10^{-6}-10^{-5}$.
Thus, far inside $r_{\rm sg}$ but much beyond $r_s$, the metric potentials have already
become small and we recover the standard Newtonian gauge
\ba
r \gg r_s: && ds^2 = - (1+2\Phi) dt^2 +  (1-2\Phi) d\vec{r}^{\,2} , \nonumber \\
&& \Phi \ll 1 , \;\; f = 1+2\Phi , \;\; h = 1-2\Phi .
\label{eq:ds2-Newtonian}
\ea
Close to the BH, where the metric potentials are governed by the Schwarschild metric
induced by the BH, we have
\be
r_s \ll r \ll r_{\rm sg} : \;\;\; \Phi= -\frac{r_s}{2r}= -\frac{{\cal G}M}{r} .
\label{eq:Phi-BH}
\ee
Far from the BH, where the contribution from the scalar
cloud to the gravitational potential becomes dominant, $\Phi$ is given by the scalar field
Poisson equation
\be
r \gg r_{\rm sg}  : \;\;\; \nabla^2 \Phi = 4\pi{\cal G} \rho_\phi ,
\label{eq:Poisson}
\ee
where $\rho_\phi$ is the scalar field energy density.
In other words, assuming a spherically symmetric scalar cloud, the metric
(\ref{eq:ds2-Schwarzschild-isotropic}) applies to all radii $r>r_s/4$.
Within radius $r_{\rm sg}$ the metric functions $f(r)$ and $h(r)$ are given by
Eq.(\ref{eq:f-h-def}), whereas beyond $r_{\rm sg}$ they are given by Eq.(\ref{eq:Poisson})
with the weak-gravity mapping (\ref{eq:ds2-Newtonian}).

In the static spherical metric (\ref{eq:ds2-Schwarzschild-isotropic}) the scalar-field
Klein-Gordon equation writes
\be
\frac{\partial^2\phi}{\partial t^2} - \sqrt{\frac{f}{h^3}} \vec\nabla \cdot ( \sqrt{f h}
\vec\nabla \phi ) + f \frac{\partial V}{\partial \phi} = 0 .
\label{eq:KG-phi-1}
\ee
This also directly follows from the action $S_\phi$ written in spherical coordinates,
\ba
&& S_\phi = \int dt dr d\theta d\varphi \sqrt{f h^3} r^2 \sin\theta \left[ \frac{1}{2 f}
\left( \frac{\partial \phi}{\partial t} \right)^2 - \frac{1}{2 h}
\left( \frac{\partial \phi}{\partial r} \right)^2 \right.
\nonumber \\
&& \left. - \frac{1}{2 h r^2} \left( \frac{\partial \phi}{\partial \theta} \right)^2
- \frac{1}{2 h r^2 \sin^2\theta} \left( \frac{\partial \phi}{\partial \varphi} \right)^2 - V(\phi) \right] .
\ea

\subsection{Eddington time coordinate}
\label{sec:Eddington}

The Schwarzschild and isotropic coordinates lead to a coordinate singularity at the
Schwarzschild radius $r_s$. As is well known, this is not a true geometrical singularity,
and one can choose coordinate systems that describe all space down to the physical
singularity at $\tilde r=0$.
For illustration, we shall consider the metric associated with the Schwarzschild radial
coordinate $\tilde r$ and the Eddington time $\tilde t$, defined by \cite{Blau-2017}
\be
\tilde{t} = t + r_s \ln \left| \frac{\tilde r}{r_s} - 1 \right| .
\label{eq:Eddington-time}
\ee
This gives the metric
\ba
ds^2 & = & - \left(1-\frac{r_s}{\tilde r} \right) d\tilde{t}^2 + 2 \frac{r_s}{\tilde r} d\tilde{t} d\tilde{r}
+ \left(1+\frac{r_s}{\tilde r}\right) d\tilde{r}^2 \nonumber \\
&& + \tilde r^2 d\vec\Omega^2 ,
\label{eq:ds2-Eddington}
\ea
which is regular over all $\tilde{r} > 0$.
These coordinates $(\tilde{t},\tilde{r})$ are directly related to the
Eddington-Finkelstein coordinates \cite{Blau-2017}.
Then, we shall check that within the metric (\ref {eq:ds2-Eddington}) the energy-momentum
tensor of the scalar field remains finite at the Schwarzschild radius, $\tilde{r}=r_s$.
In particular, in the coordinates $(\tilde{t},\tilde{r})$ and for spherically
symmetric configurations, the density defined by the time-time component of the
energy-momentum tensor reads
\be
\tilde\rho_\phi \equiv - \tilde{T}^0_0 = \frac{2-f}{2}
\left( \frac{\partial \phi}{\partial\tilde t} \right)^2
+ \frac{f}{2} \left( \frac{\partial \phi}{\partial\tilde r} \right)^2 + V ,
\label{eq:rho-Eddington}
\ee
and the partial derivatives are related by
\be
\frac{\partial \phi}{\partial\tilde t} = \frac{\partial \phi}{\partial t} , \;\;\;
\frac{\partial \phi}{\partial\tilde r} = \frac{\partial \phi}{\partial r}
\frac{1}{\sqrt{f h}} + \frac{\partial \phi}{\partial t} \left( 1 - \frac{1}{f} \right) .
\ee

\section{Free scalar field}
\label{sec:free-scalar-field}

We first consider the scalar-field inflow profile around the supermassive BH
in the free case, without self-interactions.

\subsection{Equations of motion}
\label{sec:e-o-m}

In the case of the free massive scalar field, that is, when the self-interaction vanishes,
the same decomposition of the scalar field as for the nonrelativistic case can be applied.
Thus, we can write the real scalar field $\phi$ in terms of a complex scalar field $\psi$ as
\be
\phi = \frac{1}{\sqrt{2m}} \left( e^{-imt} \psi + e^{imt} \psi^\star \right) .
\label{eq:phi-psi}
\ee
As in the nonrelativistic limit, we assume that the time derivative
of $\psi$ is much smaller than $m \psi$, that is,
\be
\dot\psi \ll m \psi ,
\label{eq:dot-psi-m}
\ee
where we note $\dot\psi=\partial\psi/\partial t$.
Thus, we focus on the large-mass limit.
Then, the scalar field action reads in terms of $\psi$ as
\ba
&& S_\psi = \int dt dr d\theta d\varphi \sqrt{f h^3} r^2 \sin\theta \left[ \frac{1}{2 f}
( i \dot\psi \psi^\star - i \psi \dot\psi^\star \right. \nonumber \\
&& + m \psi \psi^\star ) - \frac{1}{2 m h}
\frac{\partial \psi}{\partial r} \frac{\partial \psi^\star}{\partial r}
- \frac{1}{2 m h r^2} \frac{\partial \psi}{\partial \theta}
\frac{\partial \psi^\star}{\partial \theta} \nonumber \\
&& \left. - \frac{1}{2 m h r^2 \sin^2\theta}
\frac{\partial \psi}{\partial \varphi} \frac{\partial \psi^\star}{\partial \varphi}
- \frac{m}{2} \psi \psi^\star \right] .
\label{eq:S-psi}
\ea
Here we have discarded the fast oscillating terms with factors $e^{\pm 2imt}$,
which almost average to zero over a period $2\pi/m$ because of the slowly-evolving
assumption (\ref{eq:dot-psi-m}).
The action (\ref{eq:S-psi}) gives the Euler-Lagrange equation of motion
\be
i \dot\psi = - \frac{1}{2m} \sqrt{\frac{f}{h^3}} \vec\nabla \cdot ( \sqrt{f h} \vec\nabla \psi )
+ m \frac{f-1}{2} \psi .
\label{eq:dot-psi-1}
\ee
In the weak gravity regime (\ref{eq:ds2-Newtonian}), for $r \gg r_s$, we recover the usual
nonrelativistic equation,
\be
r \gg r_s : \;\;\; i \dot\psi = - \frac{\vec\nabla^2 \psi}{2m} + m \Phi \psi .
\ee

The Madelung transformation \cite{Madelung_1927},
\be
\psi = \sqrt{\frac{\rho}{m}} e^{i s} , \;\;\;
\phi = \frac{\sqrt{2\rho}}{m} \cos(m t - s) ,
\label{eq:Madelung}
\ee
maps the scalar field to an hydrodynamical picture (which breaks where $|\psi|$ vanishes
as the phase $s$ becomes ill-defined), where $\rho$ plays the role of a density and
the phase $s$ defines a curl-free velocity field through
\be
\vec{v} = \frac{\vec\nabla s}{m} .
\label{eq:v-s-def}
\ee
The scalar field action reads in terms of $\rho$ and $s$ as
\ba
&& S_{\rho,s} = \int dt dr d\theta d\varphi \sqrt{f h^3} r^2 \sin\theta \biggl \lbrace
- \frac{\rho \dot s}{m f} - \frac{1}{2 m^2 h} \nonumber \\
&& \times \left[ \frac{1}{4\rho} \left( \frac{\partial \rho}{\partial r} \right)^2
\!\! + \rho \left( \frac{\partial s}{\partial r} \right)^2 \right]
- \frac{1}{2 m^2 h r^2} \left[ \frac{1}{4\rho} \left( \frac{\partial \rho}{\partial \theta} \right)^2
\right. \nonumber \\
&& \left. + \rho \left( \frac{\partial s}{\partial \theta} \right)^2 \right]
- \frac{1}{2 m^2 h r^2 \sin^2\theta} \left[ \frac{1}{4\rho}
\left( \frac{\partial \rho}{\partial \varphi} \right)^2 \!\! + \rho \left( \frac{\partial s}{\partial \varphi} \right)^2 \right] \nonumber \\
&& + \frac{\rho}{2f} - \frac{\rho}{2} \biggl \rbrace .
\ea
In the large-mass limit, the density $\rho$ and the velocity $\vec v$ remain fixed,
while the phase $s$ grows as $m$ from Eq.(\ref{eq:v-s-def}).
Thus, formally $\rho$ is of order $m^0$ and $s$ of order $m$.
Therefore, in the large-mass limit the action simplifies to
\ba
&& S_{\rho,s} = \int dt dr d\theta d\varphi \sqrt{f h^3} r^2 \sin\theta \biggl \lbrace
- \frac{\rho \dot s}{m f} - \frac{\rho}{2 m^2 h} \left( \frac{\partial s}{\partial r} \right)^{\!\! 2}
\nonumber \\
&& - \frac{\rho}{2 m^2 h r^2} \left( \frac{\partial s}{\partial \theta} \right)^{\!\! 2}
- \frac{\rho}{2 m^2 h r^2 \sin^2\theta} \rho \left( \frac{\partial s}{\partial \varphi} \right)^{\!\! 2} \\
&& + \frac{\rho (1-f)}{2f} \biggl \rbrace ,
\label{eq:S-rho-s-m}
\ea
where we only kept the leading contributions in $m$.
This corresponds to neglecting the ``quantum pressure'' term in the Euler equation.
This is valid for small spatial density gradients,
\be
| \vec\nabla\rho | \ll m \rho .
\label{eq:density-gradient}
\ee

The Euler-Lagrange equations of motion follow from the derivatives of the action
(\ref{eq:S-rho-s-m}) with respect to $s$,
\be
\dot\rho + \sqrt{\frac{f}{h^3}} \vec\nabla \cdot \left( \sqrt{f h} \rho \frac{\vec\nabla s}{m} \right) = 0 ,
\label{eq:continuity-1}
\ee
and with respect to $\rho$,
\be
\frac{\dot s}{m} + \frac{f}{h} \frac{(\vec\nabla s)^2}{2m^2} = \frac{1-f}{2} .
\label{eq:Euler-1}
\ee
Taking the gradient of the second equation and substituting the velocity field
defined in Eq.(\ref{eq:v-s-def}) we obtain
\ba
&& \dot\rho + \sqrt{\frac{f}{h^3}} \vec\nabla \cdot ( \sqrt{f h} \rho \vec{v} ) = 0 ,
\label{eq:continuity-2}
\\
&& \dot{\vec v} + \vec\nabla \left( \frac{f}{h} \frac{{\vec v}^{\,2}}{2} \right) =
- \frac{\vec\nabla f}{2} .
\label{eq:Euler-2}
\ea
%We recognize the hydrodynamic continuity and Euler equations in a curved background,
%for a pressureless fluid.
In the weak gravity regime, $r \gg r_s$, we recover the usual Newtonian limit of the fluid equations,
\ba
r \gg r_s : && \dot\rho + \vec\nabla \cdot ( \rho \vec{v} ) = 0 ,
\label{eq:coninuity-3}
\\
&& \dot{\vec v} + ({\vec v} \cdot \vec\nabla) {\vec v} = - \vec\nabla \Phi .
\label{eq:Euler-3}
\ea
This pressureless Euler equation also corresponds to the motion of free particles in
the gravitational potential $\Phi$.

\subsection{Steady state}
\label{sec:steady}

We can look for stationary solutions of the equations of motion
(\ref{eq:continuity-1})-(\ref{eq:Euler-1}), that is, the density and the velocity fields
do not depend on time, but $s$ can have a uniform time dependence.
 This corresponds to a steady inflow of DM
from infinity into the central BH. Restricting to spherically symmetric solutions,
the continuity equation (\ref{eq:continuity-2}) gives
\be
\sqrt{f h} r^2 \rho v_r = F ,
\label{eq:flux-1}
\ee
where $F<0$ is the constant inward flux per unit solid angle,
which does not depend on the radius in a steady state.
The Euler equation (\ref{eq:Euler-2}) gives
\be
v_r = - \sqrt{\frac{h (1-f)}{f}} ,  \;\;\; \frac{\partial s}{\partial r} = m v_r ,
\label{eq:vr-r}
\ee
where we choose the boundary condition $v_r \to 0$ at $r \to \infty$ to obtain
the integration constant.
Then, there is no additional uniform time dependence for $s$ and we can choose
\be
s(r) = \int^r dr \, m v_r ,
\ee
and the complex scalar field $\psi$ is given by Eq.(\ref{eq:Madelung}).
We can check that it satisfies the equation of motion (\ref{eq:dot-psi-1}) at the leading
order in  $m$, that is, when we neglect the ``quantum pressure''.

In this large-mass limit of the free scalar field, we recover the infall of independent
massive particles, which start at rest at infinity. Their free-falling velocity does not depend
on the density because there are no self-interactions.
Then, the density is simply set by Eq.(\ref{eq:flux-1}),
that is, by the requirement of a constant flux,
\be
\rho = - \frac{F}{r^2 h \sqrt{1-f}} .
\label{eq:rho-r}
\ee
In particular, the density at the Schwarschild radius, $r=r_s/4$, is finite,
$\rho(r_s/4)=-F/r_s^2$, while the velocity $v_r$ diverges as
$-1/\sqrt{f} \sim - 1/(r-r_s/4)$.

We can now check the validity of our large scalar-mass limit.
The assumption of small time derivative (\ref{eq:dot-psi-m}) is of course satisfied
as $\psi$ does not depend on time.
We can see that the density gradient $d\rho/dr$ remains finite down to the
Schwarzschild radius. Therefore, the assumption (\ref{eq:density-gradient}) of small
density gradients is valid, as long as the Schwarzschild radius is large enough,
\be
r_s \gg m^{-1} .
\label{eq:rs-m}
\ee
Using $r_s=2{\cal G}M$, this reads as
\be
m \gg 6.7 \times 10^{-19} \left( \frac{M}{10^8 M_\odot} \right)^{-1} \, {\rm eV} .
\label{eq:m-BH}
\ee
This lower bound is somewhat larger than the lower bound (\ref{eq:m-cosmo})
associated with the growth of cosmological structures.
Thus, in this article we focus on scalar field masses in the range
$10^{-19} \ll m \lesssim 1 \, {\rm eV}$.
Our results also apply to the case of astrophysical BH, with $M \sim 1 M_{\odot}$,
if $m \gg 10^{-11} \, {\rm eV}$.

\subsection{Behavior at the Schwarzschild radius}
\label{sec:Schwarzschild-behavior}

In the regime dominated by the BH gravity, $r \ll r_{\rm sg}$, we can use
the explicit expressions of $f(r)$ and $h(r)$ of Eq.(\ref{eq:f-h-def}).
This gives for the density $\rho$ and radial velocity $v_r$,
\be
\rho = - \frac{64 F r^2}{\sqrt{r_s r} (4 r+ r_s)^3} , \;\;\;
v_r = - \sqrt{ \frac{r_s}{r^3} } \frac{(4 r + r_s)^2}{4(4r-r_s)} .
\label{eq:rho-vr-free}
\ee
Integrating $v_r$ gives the phase $s$ up to an integration constant,
\be
s = - \frac{m}{2} \sqrt{ \frac{r_s}{r} } \left[ 4 r + r_s - 4 \sqrt{r_s r} \ln \left(
\frac{ 2 \sqrt{r/r_s}+1}{2 \sqrt{r/r_s}-1} \right) \right] .
\label{eq:s-log-free}
\ee
Expanding around the Schwarzschild radius, we obtain
\ba
&& \rho = - \frac{F}{r_s^2} + \frac{3 F (4r-r_s)^2}{8 r_s^4} + \dots , \nonumber \\
&& v_r = - \frac{8 r_s}{4r-r_s} + 4 + \dots , \nonumber \\
&& s = - 2 m r_s \left[ 1 + \ln \left( \frac{4 r -r_s}{4 r_s} \right) \right] + m (4 r - r_s) + \dots
\hspace{0.8cm}
\ea
Thus, the velocity $v_r$ and the phase $s$ diverge at the Schwarzschild radius, while
the amplitude of the scalar field remains finite.
However, substituting into the expression (\ref{eq:phi-psi}) and using the
Eddington time (\ref{eq:Eddington-time}) with the Schwarzschild radial coordinate,
as in the metric (\ref{eq:ds2-Eddington}), we obtain at leading order for
$\tilde{r} \to r_s$,
\be
\phi = \sqrt{ \frac{-2F}{m^2 r_s^2} } \cos[ m ( \tilde{t} + \tilde{r} + r_s (1-\ln 4) ) ] + \dots
\label{eq:phi-free-rs}
\ee
Thus, the scalar field is well defined at the Schwarzschild radius, provided we use
regular coordinates, and as expected we recover a fully ingoing solution.
The divergence of the velocity and the phase at the Schwarzschild radius
in the Schwarzschild and isotropic metrics is due to the fact that the time $t$ is not
an appropriate coordinate at the horizon. For instance, it is well known that a massive
particle does not experience anything particular as it crosses the horizon,
which takes a finite proper time, while a distant observer that uses the time $t$
will find that the particle takes an infinite time to reach the horizon (strong redshift effect).
Then, the divergence of the phase $s$ in isotropic coordinates is required by the use
of the distant-observer time $t$. It combines with the exponential factor $e^{-imt}$
in Eq.(\ref{eq:phi-psi}) so as to give a regular expression in terms of
$({\tilde t},{\tilde r})$, once we use an appropriate time coordinate.

In a similar fashion, the energy-momentum tensor associated with the
Schwarzschild or isotropic metrics diverges at the Schwarzschild radius, but
the one associated with the Eddington metric (\ref{eq:ds2-Eddington}) remains finite.

\subsection{Density profile}
\label{sec:density-free}

From Eqs.(\ref{eq:rho-Eddington}) and (\ref{eq:Madelung}), the energy density
associated with the Eddington coordinates is given, at leading order in the large-$m$
limit, by
\ba
\tilde\rho_\phi & = & \rho \biggl \lbrace \sin^2(mt-s) \left[ 2-f + \frac{1}{f} (1-f-\sqrt{1-f})^2 \right]
\nonumber \\
&& + \cos^2(mt-s) \biggl \rbrace .
\label{eq:T00-free-sin}
\ea
In terms of the flux $F$, we obtain using Eq.(\ref{eq:rho-r})
\be
\langle \tilde\rho_\phi \rangle = - \frac{F}{r_s^2} \frac{r_s^2}{2 r^2 h \sqrt{1-f}}
\left[ 3 - f + \frac{1}{f} (1-f-\sqrt{1-f})^2 \right]
\label{eq:T00-free-F}
\ee
where we took the average over the fast oscillations of period $2\pi/m$.
As expected, this scalar-field energy density remains finite at the Schwarzschild radius,
with
\be
\tilde{r} = r_s, \;\; r= \frac{r_s}{4} : \;\;\; \langle \tilde\rho_\phi \rangle = - \frac{3 F}{2 r_s^2} .
\label{eq:T00-free}
\ee
At large radii, which are still dominated by the BH gravitational potential, this gives
\be
r_s \ll r \ll r_{\rm sg}: \;\;\; \langle \tilde\rho_\phi \rangle \propto r^{-3/2} \;\;\;
\mbox{and} \;\;\; v_r \propto r^{-1/2} .
\label{eq:rho-r-free}
\ee
The scaling $v_r \propto r^{-1/2}$ corresponds to the free fall from rest at infinity,
which also gives $v_r^2 \sim \Phi \sim {\cal G} M/r$. The requirement of constant flux
through spherical shells then implies $\rho_\phi \propto r^{-3/2}$.
The density $\rho_\phi$ grows linearly with $|F|$,
as there are no self-interactions (and we neglect self-gravity near the BH).

The unit ``velocity'' obtained in the ingoing wave (\ref{eq:phi-free-rs}),
or of order unity in Eq.(\ref{eq:T00-free}) if we define an effective velocity by
$F=\langle \tilde \rho_\phi \rangle r^2 \tilde{v}_r^{\rm eff}$, shows that as expected the scalar-field
dynamics are strongly relativistic at the Schwarzschild radius.
In particular, the phase $s$ is not small and the exponent $e^{is}$ of the wave function
$\psi$ cannot be expanded over, as it must precisely combine with the factor
$e^{-imt}$ to give the regular solution (\ref{eq:phi-free-rs}).
Also, whereas $\rho$ given by Eq.(\ref{eq:rho-vr-free}) remains finite
at the Schwarzschild radius, $s$ given by Eq.(\ref{eq:s-log-free}) diverges.
This means that whereas density gradients remain small, as compared with the
scalar mass, as long as the bound (\ref{eq:rs-m}) is fulfilled,
the radial derivatives of the phase $s$ and of the wave functions $\psi$ and $\phi$
are not small and even diverge at the Schwarzschild radius.
Again, this means that one cannot use a perturbative approach in the scalar field,
even in the large scalar mass limit. One must keep the nonlinearities of the scalar field phase.

\section{Quartic interaction}
\label{sec:quartic}

We now consider the scalar-field inflow profile around the supermassive BH in the case of
quartic self-interactions (\ref{eq:V-quartic}).

\subsection{Large-mass approximation}
\label{sec:large-mass}

For spherical modes and the quartic self-interaction (\ref{eq:V-quartic}) the
nonlinear Klein-Gordon equation (\ref{eq:KG-phi-1}) reads
\ba
&& \frac{\partial^2\phi}{\partial t^2} - \sqrt{\frac{f}{h^3}} \frac{1}{r^2} \frac{\partial}{\partial r}
\left[ \sqrt{f h} r^2 \frac{\partial\phi}{\partial r} \right] + f m^2 \phi + f \lambda_4 \phi^3 = 0 .
\nonumber \\
&&
\label{eq:KG-r}
\ea

If we discard the radial derivatives we recognize the standard Duffing equation,
which describes a nonlinear oscillator with a cubic nonlinearity \cite{Kovacic-2011}.
Its solution can be written as $\phi_0 {\rm cn}(\omega t - \beta,k)$, where
${\rm cn}(u,k)$ is the Jacobi elliptic function \cite{Gradshteyn1965,Byrd-1971}
of argument $u$ and modulus $k$.
The angular frequency $\omega$ and the modulus $k$ are functions of the
amplitude $\phi_0$, as for anharmonic oscillators the frequency depends on the
amplitude of the oscillations.
The harmonic case $\lambda_4=0$ corresponds to $k=0$ as
${\rm cn}(u,0)=\cos(u)$.
For general $k$, the Jacobi elliptic function ${\rm cn}(u,k)$ is a periodic function
of $u$ with period $4 {\bf K}$, where ${\bf K}(k)$ is the complete elliptic integral
of the first kind, defined by \cite{Gradshteyn1965,Byrd-1971}
\be
0 \leq k < 1 : \;\;\; {\bf K}(k) = \int_0^{\pi/2} \frac{d\theta}{\sqrt{1-k^2\sin^2\theta}} ,
\ee
and ${\bf K}(0)=\pi/2$.

Taking into account the radial dependence, we can look for a solution of the form
\be
\phi = \phi_0(r) \, {\rm cn}[ \omega(r) t - {\bf K}(r) \beta(r), k(r) ] ,
\label{eq:phi-cn-def}
\ee
where we noted ${\bf K}(r) \equiv {\bf K}[k(r)]$.
This is understood as the leading-order approximation in the limit $m \to \infty$,
where spatial gradients of the amplitude $\phi_0$ and the modulus $k$
are much below $m$, while both $\omega$ and $\beta$
are of order $m$.
The amplitude $\phi_0$, the angular frequency $\omega$, the
phase $\beta$ and the modulus $k$ are slow functions of the radius.
Thus, Eq.(\ref{eq:phi-cn-def}) is a generalization of the free-scalar solution
(\ref{eq:Madelung}) to the case of nonzero quartic self-interaction,
in the same large-mass approximation.
To ensure that spatial gradients do not increase with time, all radii must
oscillate in phase, with a common period $T=2\pi/\omega_0$.
Thus, $\omega T=4{\bf K}$ and the angular frequency $\omega(r)$ is set by the modulus
$k(r)$ as
\be
\omega(r) = \frac{2{\bf K}(r)}{\pi} \omega_0 ,
\label{eq:omega-omega0}
\ee
where $\omega_0$ is a parameter to be determined.
The synchronous oscillation can be seen from the series expansion
of the Jacobi elliptic function, which gives \cite{Gradshteyn1965,Byrd-1971}
\be
\phi = \phi_0 \frac{2\pi}{k{\bf K}} \sum_{n=0}^{\infty} \frac{q^{n+1/2}}{1+q^{2n+1}}
\cos [ (2n+1) (\omega_0 t - \pi \beta/2) ] ,
\label{eq:cn-series}
\ee
with $q=e^{-\pi {\bf K'}/{\bf K}}$, where ${\bf K'}={\bf K}(k')$ with $k'=\sqrt{1-k^2}$.

From Eq.(\ref{eq:phi-cn-def}) the time derivative is
\be
\frac{\partial\phi}{\partial t} = \phi_0 \omega \frac{\partial {\rm cn}}{\partial u} .
\label{eq:dphi-dt-cn}
\ee
At leading order in the large-$m$ limit, the radial derivative reads from
Eq.(\ref{eq:cn-series}) as
\be
\frac{\partial\phi}{\partial r} = - \phi_0 {\bf K} \beta' \frac{\partial {\rm cn}}{\partial u} + \dots ,
\label{eq:dphi-dr-cn}
\ee
where the dots stand for subleading terms, as we assume that the phase $\beta$
is formally of order $m$. Here $\beta' = d\beta/dr$.
Substituting into the nonlinear Klein-Gordon equation (\ref{eq:KG-r}) gives
\be
\phi_0 \left[ \omega^2 - \frac{f}{h} ( {\bf K} \beta' )^2 \right] \frac{\partial^2 {\rm cn}}{\partial u^2}
+ f m^2 \phi_0 {\rm cn} + f \lambda_4 \phi_0^3 {\rm cn}^3 = 0 ,
\label{eq:KG-cn}
\ee
where we only kept the term of order $m^2$ among the radial derivative contributions.
Thus, at this order, we can see that the radial derivatives do not change the structure
of Eq.(\ref{eq:KG-cn}). This is why it again admits the Jacobi elliptic function as a solution.
Thus, using the property
\be
\frac{\partial^2 {\rm cn}}{\partial u^2} = (2k^2-1) {\rm cn} - 2 k^2 {\rm cn}^3 ,
\label{eq:d2cndu2}
\ee
the Klein-Gordon equation (\ref{eq:KG-cn}) is satisfied as soon as the coefficients
of the factors ${\rm cn}$ and ${\rm cn}^3$ vanish.
This gives the two conditions
\ba
&& \frac{\pi^2 f}{4 h} \beta'^2 = \omega_0^2 - \frac{f m^2 \pi^2}{(1-2k^2) 4 {\bf K}^2} ,
\label{eq:beta-1}
\\
&& \frac{\lambda_4 \phi_0^2}{m^2} = \frac{2k^2}{1-2k^2} .
\label{eq:lambda4-k}
\ea
We recover in Eq.(\ref{eq:lambda4-k}) that the free scalar case, $\lambda_4=0$,
corresponds to $k=0$.
Equation (\ref{eq:beta-1}) is the generalization of the Euler equation (\ref{eq:Euler-1}),
$\pi \beta'/(2m)$ plays the role of the radial velocity $v_r=m^{-1} ds/dr$ and
$\pi \beta/2$ plays the role of the phase $s$.

\subsection{Boundary conditions}
\label{sec:boundary-all}

\subsubsection{Large-radius boundary condition}
\label{sec:large-radius-boundary}

At large radii, $r \gg r_{\rm sg}$, the gravitational field is small and set by the self-gravity
of the scalar cloud.
Therefore, we match the solution (\ref{eq:phi-cn-def}) to the soliton profile obtained
for the self-gravitational nonrelativistic scalar cloud \cite{Brax:2019fzb}.

\paragraph{Scalar-field soliton}
\label{sec:soliton}

In this regime, we can decompose the scalar field $\phi$ as in Eq.(\ref{eq:phi-psi})
and use the Madelung transformation (\ref{eq:Madelung}) for the complex field $\psi$.
Taking into account the quartic self-interaction, which is subdominant with respect
to the quadratic potential $m^2\phi^2/2$, the continuity equations
(\ref{eq:continuity-1}) and (\ref{eq:continuity-2}) take again the usual form
(\ref{eq:coninuity-3}),
\be
\dot\rho + \frac{\vec\nabla \cdot ( \rho \vec\nabla s )}{m} = 0 , \;\;\;
\dot\rho + \vec\nabla \cdot ( \rho \vec{v} ) = 0 ,
\ee
whereas the Euler equations (\ref{eq:Euler-1}) and (\ref{eq:Euler-2}) become
\ba
&& \frac{\dot s}{m} + \frac{(\vec\nabla s)^2}{2m^2} = - (\Phi+\Phi_{\rm I}) , \nonumber \\
&& \dot{\vec v} + ( \vec{v} \cdot \vec\nabla ) \vec v = - \vec\nabla (\Phi+\Phi_{\rm I}) ,
\label{eq:Euler-4}
\ea
where $\Phi_{\rm I}$ is given by \cite{Brax:2019fzb}
\be
\Phi_{\rm I}(\rho) = \frac{\rho}{\rho_a} , \;\;\; \rho_a \equiv \frac{4 m^4}{3\lambda_4} .
\label{eq:Phi-I-def}
\ee
This ``pressure'' associated with the self-interaction $\Phi_{\rm I}$ allows the scalar cloud
to reach an hydrostatic equilibrium, where this repulsive self-interaction balances the
self-gravity.
This gives the soliton profile \cite{Brax:2019fzb}
\be
\rho(r) = \rho_s(0) \frac{\sin(r/r_a)}{r/r_a} , \;\;\;
\Phi_{\rm I}(r) = \Phi_{{\rm I}s}(0) \frac{\sin(r/r_a)}{r/r_a} ,
\ee
with ${\vec v} = 0$ and
\be
r_a = \frac{1}{\sqrt{4\pi{\cal G} \rho_a}} .
\label{eq:r_a-rho_a}
\ee
The soliton has a flat inner core and a finite radius $R_s=\pi r_a$.
Inside the soliton, the hydrostatic equilibrium condition (\ref{eq:Euler-4})
gives $\vec\nabla (\Phi+\Phi_{\rm I}) =0$, and we have
\be
r \leq R_s : \;\;\; \Phi+\Phi_{\rm I} = \alpha ,
\label{eq:alpha-def}
\ee
where $\alpha$ is a constant, given by the value of the Newtonian potential
at the boundary of the soliton,
\be
\alpha = \Phi(R_s) ,
\label{eq:alpha-Phi-Rs}
\ee
as $\Phi_{\rm I}(R_s)=0$.
In terms of the scalar fields $\psi$ and $\phi$ this gives
\be
\psi = \sqrt{\frac{\rho}{m}} e^{-i\alpha m t} , \;\;\; \mbox{hence} \;\;\;s= - \alpha m t,
\ee
and
\be
\phi = \frac{\sqrt{2\rho}}{m} \cos[ (1+\alpha) m t ] .
\label{eq:phi-soliton}
\ee

\paragraph{Large-radius solution}
\label{sec:large-radius-solution}

At large radii but within the soliton radius, $r_{\rm sg} \ll r \ll R_s$, we are in the weak-gravity
regime and we approach the soliton core solution,
with $\Phi \simeq \Phi_s(0) \lesssim 10^{-5}$ and $\rho \simeq \rho_s(0)$.
We also have $\Phi_{\rm I} = \alpha - \Phi \simeq -\Phi_s(0)$, and the self-interaction
potential $V_{\rm I} \sim \rho \Phi_{\rm I} \ll \rho$ is much smaller than the quadratic part,
hence $\lambda_4 \phi^4 \ll m^2 \phi^2$. Therefore, we can see from Eq.(\ref{eq:lambda4-k})
that we have at leading order
\be
k^2 = \frac{\lambda_4 \phi_0^2}{2m^2} + \dots \ll 1 ,
\label{eq:k2-small}
\ee
where the dots stand for higher-order terms.
From the expansion (\ref{eq:cn-series}) and the series expansions \cite{Byrd-1971}
\ba
&& {\bf K}(k) = \frac{\pi}{2} \left( 1 + \frac{k^2}{4} + \dots \right) , \\
&& q(k) = \frac{k^2}{16} \left( 1 + \frac{k^2}{8} + \dots \right) ,
\ea
we obtain at leading order
\be
k \ll 1: \;\;\; \phi = \phi_0 \cos( \omega_0 t - \pi \beta/2) + \dots
\ee
The comparison with Eq.(\ref{eq:phi-soliton}) gives
\be
r_{\rm sg} \ll r \ll R_s : \;\;\; \phi_0(r) = \frac{\sqrt{2\rho_s(0)}}{m}  , \;\;\; \beta \simeq 0 ,
\label{eq:phi0-rhos}
\ee
and
\be
\omega_0 = (1+\alpha) m .
\label{eq:omega0-alpha}
\ee
Indeed, as the soliton solution (\ref{eq:phi-soliton}) corresponds to hydrostatic equilibrium
with $\vec v=0$, the ``velocity'' $\beta$ must become negligible at large radii in order
to match with the soliton.
We can now check that this is consistent with Eqs.(\ref{eq:beta-1})-(\ref{eq:lambda4-k}).
Eq.(\ref{eq:beta-1}) with $\beta=0$ gives, at leading order in $\Phi$ and $k^2$,
\be
\omega_0 = m \left(1+\Phi+\frac{3}{4} k^2 \right) .
\label{eq:omega0-Phi-k2}
\ee
On the other hand, Eq.(\ref{eq:lambda4-k}) gave Eq.(\ref{eq:k2-small}).
Using Eq.(\ref{eq:phi0-rhos}) this yields
\be
k^2 = \frac{\lambda_4 \rho}{m^4} = \frac{4}{3} \Phi_{\rm I} .
\label{eq:k2-rho-Phi-I}
\ee
Then, Eq.(\ref{eq:omega0-Phi-k2}) reads
$\omega_0 = m (1+\Phi+\Phi_{\rm I}) =  m (1+\alpha)$, where we used the hydrostatic
result (\ref{eq:alpha-def}), and we recover Eq.(\ref{eq:omega0-alpha}).
This shows this large-radius asymptote is self-consistent, provided $\beta$ is negligible.
This gives the large-radius asymptotic values of $\phi_0(r)$ and $k(r)$,
from Eqs.(\ref{eq:phi0-rhos}) and (\ref{eq:k2-rho-Phi-I}), in the constant-density core
of the soliton.
The uniform oscillation frequency $\omega_0$ is then set by this large-radius boundary
condition in Eq.(\ref{eq:omega0-alpha}).
Note that typically $\alpha \lesssim 10^{-5}$ from Eq.(\ref{eq:alpha-Phi-Rs}).
Thus, the angular oscillation frequency $\omega_0$ remains very close to $m$.

\subsubsection{Small-radius boundary condition}
\label{sec:small-radius}

Close to the Schwarzschild radius, we can expect the self-interaction ``pressure'' to be
negligible and to recover the free-scalar infall (\ref{eq:vr-r})
(but we shall see below that the self-interaction plays a role for the scalar profile
down to the Schwarzschild radius, as it dictates the constant flux of the steady state).
Indeed, as long as $k$ remains below $1/\sqrt{2}$, the last term in the generalized
Euler equation (\ref{eq:beta-1}) becomes negligible as $f \to 0$ at the Schwarzschild radius,
and Eq.(\ref{eq:beta-1}) gives
\be
r \to \frac{r_s}{4} : \;\;\; \frac{\pi}{2} \beta' = - \omega_0 \sqrt{\frac{h}{f}} .
\label{eq:betap-small-radius}
\ee
This agrees indeed with Eq.(\ref{eq:vr-r}) (except for the prefactor $\alpha$ associated
with the finite soliton size).

\subsection{Steady state and constant flux}
\label{sec:steady-quartic}

So far, any profile $k(r)$ with the outer boundary condition (\ref{eq:k2-rho-Phi-I})
and $k(r)<1/\sqrt{2}$ at all radii provides a leading-order solution (\ref{eq:phi-cn-def}).
Indeed, given $k(r)$, Eq.(\ref{eq:beta-1}) provides the ``velocity'' $\beta'$ while
Eq.(\ref{eq:lambda4-k}) provides the amplitude $\phi_0$, i.e. the ``density''.
Clearly, we do not expect such a large space of physical solutions. It would seem more
natural to recover a specific profile, such as the unique transonic solution found
for hydrodynamics in nonrelativistic \cite{Bondi:1952ni} and relativistic \cite{Michel:1972} infall.
In fact, at this stage we miss a constant flux constraint associated with a continuity equation,
as in Eq.(\ref{eq:flux-1}).
In the relativistic case, the continuity equation is associated with the component
$\nu=0$ of the conservation equations $\nabla_\mu T^\mu_\nu = 0$.
The energy-momentum tensor of the scalar field $\phi$ gives
\be
\rho_\phi \equiv - T^0_0 = \frac{1}{2f} \left( \frac{\partial\phi}{\partial t} \right)^2
+ \frac{1}{2h} \left( \frac{\partial\phi}{\partial r} \right)^2 + V
\ee
and
\be
T^r_0 = \frac{1}{h} \frac{\partial\phi}{\partial r} \frac{\partial\phi}{\partial t} .
\ee
At leading order in the large-mass limit, we obtain from Eqs.(\ref{eq:phi-cn-def}),
(\ref{eq:dphi-dt-cn})-(\ref{eq:dphi-dr-cn}), and (\ref{eq:beta-1})-(\ref{eq:lambda4-k}),
\ba
\rho_\phi & = & \frac{(1-k^2)m^2\phi_0^2}{2(1-2k^2)} + \phi_0^2 \frac{({\bf K}\beta')^2}{h}
[ 1 - k^2 + (2k^2-1) {\rm cn}^2 \nonumber \\
&& - k^2 {\rm cn}^4 ]  ,
\label{eq:rho-T00}
\ea
and
\be
T^r_0 = - \phi_0^2 \omega \frac{{\bf K}\beta'}{h} \left( \frac{\partial {\rm cn}}{\partial u} \right)^2 .
\ee
Then, using again Eqs.(\ref{eq:dphi-dt-cn})-(\ref{eq:dphi-dr-cn}) and
(\ref{eq:d2cndu2}), we can check that the conservation equation $\nabla_\mu T^\mu_0 = 0$,
which reads,
\be
\dot\rho - \frac{1}{\sqrt{f h^3} r^2} \frac{\partial}{\partial r} \left[ \sqrt{f h^3} r^2 T^r_0 \right] = 0 ,
\label{eq:continuity-rho-Tr0}
\ee
is satisfied at the leading order.
We can note that $\rho_\phi$ is not constant with time, as the terms ${\rm cn}^2$
and ${\rm cn}^4$ in the bracket in Eq.(\ref{eq:rho-T00}) oscillate with the frequency
$\omega_0$.
At the leading order, the continuity equation (\ref{eq:continuity-rho-Tr0}) is governed
by the fast oscillation of these terms.
However, to ensure that subleading orders do not show secular terms that grow with time,
we clearly require that in the steady state the averaged value of $\rho_\phi$ over one oscillation
period does not depend on time.
This gives the condition of constant flux
\be
F = - \sqrt{f h^3} r^2 \langle T^r_0 \rangle = \sqrt{f h} r^2 \phi_0^2 \omega {\bf K} \beta'
\langle \left( \frac{\partial {\rm cn}}{\partial u} \right)^2 \rangle ,
\label{eq:F-Tr0-betap}
\ee
where $\langle \dots \rangle$ denotes the average over one oscillation period $T=2\pi/\omega_0$.
Using Eqs.(\ref{eq:omega-omega0}), (\ref{eq:beta-1})-(\ref{eq:lambda4-k}),
and (\ref{eq:omega0-alpha}), we can write the flux in terms of $k(r)$,
\ba
F & = & F_s x^2 h
 \left( \frac{2 {\bf K}}{\pi} \right)^2
\langle \left( \frac{\partial {\rm cn}}{\partial u} \right)^2 \rangle
\frac{2 k^2}{1-2 k^2}
 \nonumber \\
&& \times \sqrt{ 1 - \frac{\pi^2 f}{(1+\alpha)^2 4 {\bf K}^2 (1-2k^2)} }  ,
\label{eq:Flux-Fs}
\ea
where we defined the dimensionless radial coordinate
\be
x = \frac{r}{r_s} > \frac{1}{4} ,
\ee
and the characteristic flux
\be
F_s = - \frac{r_s^2 m^4 (1+\alpha)^2}{\lambda_4} \simeq - \frac{r_s^2 m^4}{\lambda_4} ,
\label{eq:Fs-def}
\ee
as typically $\alpha \lesssim 10^{-5}$.
The average value of $(\frac{\partial {\rm cn}}{{\partial u}})^2$ is
\be
\langle \left( \frac{\partial {\rm cn}}{\partial u} \right)^2 \rangle = 1 - k^2 + (2k^2-1) C_2
- k^2 C_4
\ee
with \cite{Kovacic-2011}
\ba
&& C_2 \equiv \langle {\rm cn}^2 \rangle = \frac{1}{k^2} \left( \frac{\bf E}{\bf K} + k^2 -1 \right) , \\
&& C_4 \equiv \langle {\rm cn}^4 \rangle = \frac{1}{3k^2} ( 2 (2k^2-1) C_2 + 1 -k^2 ) ,
\ea
where ${\bf E}(k)$ is the complete elliptic integral of the second kind,
defined by \cite{Gradshteyn1965,Byrd-1971}
\be
0 \leq k < 1 : \;\;\; {\bf E}(k) = \int_0^{\pi/2} d\theta \, \sqrt{1-k^2\sin^2\theta}  .
\ee

We can see from Eq.(\ref{eq:Fs-def}) that the flux diverges as $1/\lambda_4$.
This is not surprising, since for vanishing self-interaction we must recover the free-scalar
case studied in sec.~\ref{sec:free-scalar-field}, where the flux is arbitrary and has
no upper bound.
We also find that the flux scales as $r_s^2 m^4$, which is also natural,
as we can expect $F \sim \rho r^2 v_r$, with $r = r_s/4$, $v_r \sim 1$
at the Schwarzschild radius and $\rho \sim m^4$ from dimensional analysis.

\subsection{Critical solution}
\label{sec:critical}

\begin{figure}
\begin{center}
\epsfxsize=8.8 cm \epsfysize=6 cm {\epsfbox{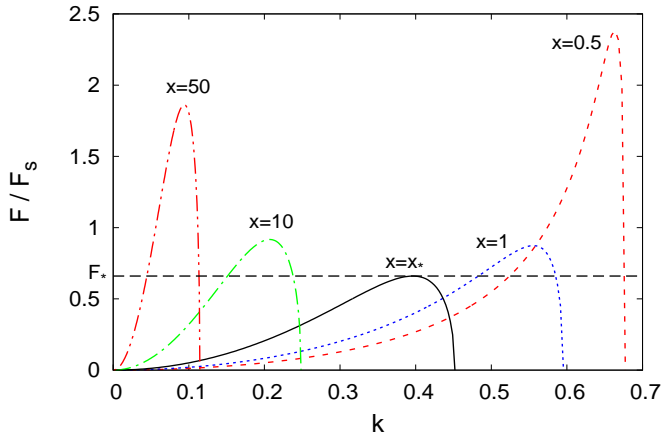}}
\end{center}
\caption{Normalized flux $F(k,x)/F_s$ as a function of the modulus $k$, for various values
of the radial coordinate $x$, from Eq.(\ref{eq:Flux-Fs}).
The horizontal dotted line is the minimum value $F_\star \simeq 0.66$ of the peak,
reached for $x=x_\star\simeq 2.43$.}
\label{fig_flux}
\end{figure}

\begin{figure}
\begin{center}
\epsfxsize=8.8 cm \epsfysize=6 cm {\epsfbox{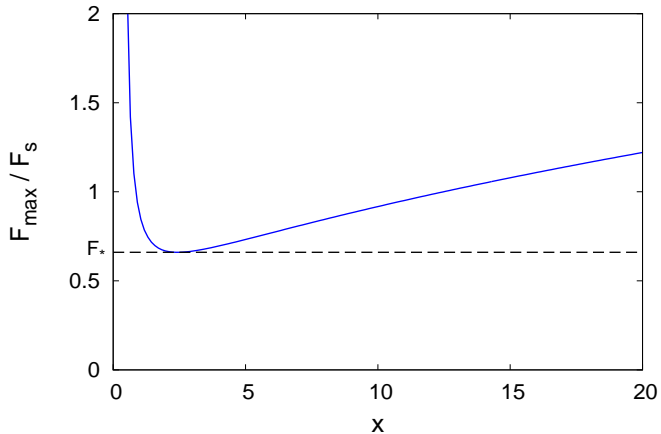}}
\end{center}
\caption{Peak value $F_{\max}(x)/F_s$ as a function of the radial coordinate $x$.
The horizontal dotted line is the minimum value $F_\star \simeq 0.66$.}
\label{fig_Fmax}
\end{figure}

\subsubsection{Function $F(k,x)$}
\label{sec:function-F}

For each radius $x$, Eq.(\ref{eq:Flux-Fs}) gives the flux $F$ as a function of $k$.
We show in Fig.~\ref{fig_flux} the normalized flux $F/F_s$ as a function of the modulus $k$
for several values of the radial coordinate $x$.
The modulus $k$ is constrained to range between 0 and the value $k_+(x)<1/\sqrt{2}$
where the square root vanishes.
The flux vanishes at both boundaries, $k=0$ and $k=k_+$, and shows a single
maximum $| F_{\rm max}(x) |$ at a position $k_{\max}(x)$ somewhat below $k_+(x)$.
The upper bound $k_+$ and the peak at $k_{\max}$ shift to lower values as $x$ grows.
The maximum $| F_{\rm max}(x) |$ grows at both small and large $x$, and shows a minimum
at $x_\star\simeq 2.43$ with
\be
F_c \equiv F_{\rm max}(x_\star) = F_\star F_s \;\;\; \mbox{with} \;\;\; F_\star \simeq 0.66  .
\ee
We show $F_{\rm max}(x)/F_s$ in Fig.~\ref{fig_Fmax}.
In Figs.~\ref{fig_flux} and \ref{fig_Fmax}, we use for the metric functions $h(x)$ and $f(x)$
the Schwarzschild functions (\ref{eq:f-h-def}). At the transition radius $r_{\rm sg}$, the gravitational
potential receives equal contributions from the central BH and the scalar cloud,
and at larger radius inside the soliton core it remains almost constant, equal to the soliton
core value $\Phi_s(0)$. Therefore, beyond $r_{\rm sg}$ the factors $h$ and $f$ are almost constant
and the flux function $F(x,k)$ keeps a constant shape in $k$, with a simple multiplicative
factor $x^2$. Thus, beyond $r_{\rm sg}$ the peak value $| F_{\max}(x) |$ keeps increasing,
as $x^2$.

\begin{figure}
\begin{center}
\epsfxsize=8.8 cm \epsfysize=6 cm {\epsfbox{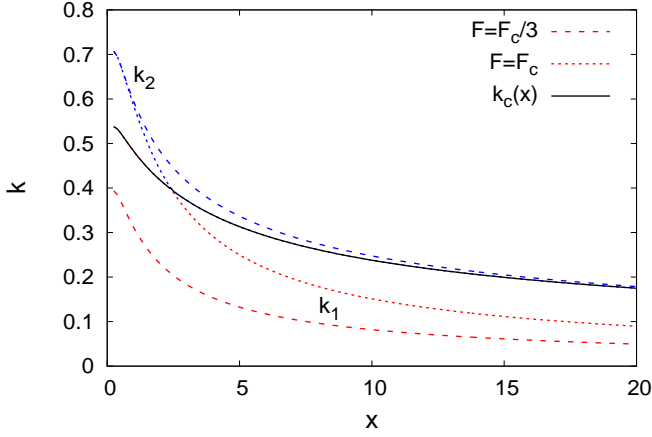}}
\end{center}
\caption{Moduli $k_1(x)$ and $k_2(x)$ for a constant flux $F_c/3$ (dashed lines)
and $F_c$ (dotted lines). The critical modulus $k_c(x)$ (solid line) is equal to $k_1$
for $x<x_\star$ and to $k_2$ for $x>x_\star$, with $F=F_c$.}
\label{fig_k2_k2}
\end{figure}

This behavior of $F(k,x)$ selects a unique value for the flux, in a fashion similar to the
unique transonic solution found in the case of hydrodynamical infall onto a BH
\cite{Bondi:1952ni,Michel:1972}.
First, we can see that $|F|$ must be smaller or equal to the critical value $|F_c|$,
otherwise there would exist no solution $k(x)$ to the flux constraint equation
(\ref{eq:Flux-Fs}) around $x_\star$.
If $|F|<|F_c|$ there exist two distinct solutions $k_1(x)<k_2(x)$ at each radius,
on either side of the peak $k_{\max}(x)$, and a continuous function $k(x)$
must remain on the same side of the peak throughout.
It is only for the critical value $F=F_c$ that the function $k(x)$ can
switch from the branch $k_1(x)$ to $k_2(x)$, at the radius $x_\star$ where both
solutions coincide with the peak.
The two solutions $k_1(x)<k_2(x)$ are shown in Fig.~\ref{fig_k2_k2}
for $F=F_c/3$ (the upper and lower dashed curves that do not meet) and
for $F=F_c$ (the inner dotted curves that meet at $x_\star\simeq 2.43$,
which coincide with the critical solution $k_c(x)$, shown by the solid line, on either
side of $x_\star$).

As we shall see below, the boundary conditions require that $k=k_2(x)$ at large
radii and $k=k_1(x)$ close to the Schwarzschild radius.
Therefore, the function $k(x)$ must change branches at some intermediate radius.
This selects the flux $F=F_c$ as the only physical value
and the solution $k_c(x)$ that switches from $k_1$ to $k_2$,
as shown by the solid line in Fig.~\ref{fig_k2_k2}.
This is similar to the hydrodynamical case \cite{Bondi:1952ni,Michel:1972},
which selects the only value of the flux that provides a transonic solution that connects
the subsonic (i.e. low velocity) branch at large radii with the supersonic (i.e. high velocity)
branch at low radii.

\subsubsection{Boundary conditions}
\label{sec:boundary}

To obtain the boundary condition at large radius for the modulus $k(x)$,
we consider the behavior of $F(k,x)$ at small $k$.
Indeed, as seen in Fig.~\ref{fig_flux}, at large radii the upper boundary $k_+(x)$
becomes much smaller than unity.
Then, Eq.(\ref{eq:Flux-Fs}) gives at leading order
\be
k \ll 1 : \;\;\; \frac{F}{F_s}= x^2 h k^2 \sqrt{ 1 - \frac{f (1+3 k^2/2)}{(1+\alpha)^2} } .
\label{eq:k-small-F}
\ee
At large radii inside the soliton core, we obtain
\be
r \gg r_{\rm sg} , \;\; k \ll 1 : \;\; \frac{F}{F_s} = x^2 k^2
\sqrt{ \frac{3}{2} \left( \frac{\lambda_4\rho}{m^4} - k^2 \right) }
\label{eq:F-k-small-r-large}
\ee
at lowest order in $k^2$, $\Phi$ and $\alpha$, and we used Eqs.(\ref{eq:Phi-I-def})
and (\ref{eq:alpha-def}).
Since at these radii $\lambda_4 \rho/m^4 \ll 1$, the small-$k$ expansion is valid up to
$k_+$, which is thus given by
\be
r \gg r_{\rm sg} : \;\; k_+ = \sqrt{\frac{\lambda_4\rho}{m^4}} \ll 1 .
\ee
We can see from Eq.(\ref{eq:k2-rho-Phi-I}) that the large-radius boundary condition
is in fact $k(r) = k_+$, when we neglect the velocity as in the analysis of
sec.~\ref{sec:large-radius-solution}.
In agreement with Eq.(\ref{eq:F-k-small-r-large}), we find that this boundary condition
with a zero velocity implies a zero flux $F$.
In practice, the matching to the static soliton is not perfect and there remains a nonzero
velocity $\beta'$, associated with a nonzero flux $F$.
This approximate matching is meaningful as long as the velocity at the outer boundary
of the core, $r \sim R_s/10$, is sufficiently small. In other words, it must be much smaller
than the free-fall velocity at that radius, and the mass loss onto the BH
should remain much smaller than the soliton mass over the time of interest.
We shall check below in section~\ref{sec:lifetime} that this is indeed the case.
This also means that at large radii the modulus $k(x)$ must be on the upper branch
$k_2(x)$, close to the upper boundary $k_+(x)$,
\be
x \gg r_{\rm sg}/r_s : \;\;\; k(x) = k_2(x) .
\label{eq:k2}
\ee

At the Schwarzschild radius, $x \to 1/4$, $h(x) \to 16$ and $f(x) \to 0$.
Therefore, the square root in Eq.(\ref{eq:Flux-Fs}) goes to unity (unless $k \to 1/\sqrt{2}$).
More physically, the square root comes from the ``velocity'' factor $\beta'$ of
Eq.(\ref{eq:F-Tr0-betap}), through Eq.(\ref{eq:beta-1}). Close to the Schwarzschild radius,
the velocity should be large and close to unity, as found in Eq.(\ref{eq:betap-small-radius}),
and the self-interaction become negligible as we recover the free fall onto the BH.
This means that the square root in Eq.(\ref{eq:Flux-Fs}) goes to unity. Then,
the small value of the flux $F$ as compared with the local peak value $F_{\max}(x)$
is reached by having a small value of $k$, thanks to the prefactor $k^2$,
rather than by having a large value of $k$ close to the upper boundary $k_+$
where the square root vanishes.
This means that at small radii the modulus $k(x)$ must be on the lower branch
$k_1(x)$, close to zero,
\be
x \simeq 1/4 : \;\;\; k(x) = k_1(x) .
\label{eq:k1}
\ee
Thus, as announced above, the boundary conditions (\ref{eq:k2})-(\ref{eq:k1})
imply that the physical solution $k(x)$ must change from the upper to the lower branch,
as we get closer to the BH.
As explained in sec.~\ref{sec:function-F}, this selects the unique value $F_c$
for the flux and a unique function $k(x)$.

\subsubsection{Critical solution}
\label{sec:critical-k}

Thus, the unique function $k_c(x)$, shown by the solid line in Fig.~\ref{fig_k2_k2},
verifies
\ba
F=F_c , \;\;\; && k_c(x) = k_1(x) \;\; \mbox{for} \;\; x<x_\star , \nonumber \\
&& k_c(x) = k_2(x) \;\; \mbox{for} \;\; x>x_\star .
\ea
At the Schwarzschild radius we obtain
\be
r= r_s/4 : \;\;\; k_c(1/4) \equiv k_s \simeq 0.54 ,
\label{eq:ks-def}
\ee
while $k_c(x)$ decreases at large radius.
From Eq.(\ref{eq:F-k-small-r-large}), with $F=F_c$, we obtain at large radii
\be
r \gg r_{\rm sg} : \;\;\; k_c(x)^2 = \frac{\lambda_4\rho}{m^4} - \frac{2}{3x^4}
\left( \frac{F_\star m^4}{\lambda_4\rho} \right)^2 .
\ee
Eq.(\ref{eq:beta-1}) gives
\be
v_r \equiv \frac{\pi \beta'}{2m} = - \sqrt{ \frac{h}{f} }
\sqrt{ (1+\alpha)^2 - \frac{\pi^2 f}{(1-2k^2) 4 {\bf K}^2} }
\label{eq:vr-sqrt}
\ee
where we made the identification $v_r = \pi\beta'/2m$, which holds in the weak gravity
nonrelativistic limit, as explained below Eq.(\ref{eq:lambda4-k}).
The density $\rho_\phi$ defined in Eq.(\ref{eq:rho-T00}) reads
\ba
\frac{\langle \rho_\phi \rangle}{\rho_a} & = & \frac{3 k^2}{4 (1-2k^2)}
\biggl [ \frac{1-k^2}{1-2k^2} + \frac{8 {\bf K}^2 v_r^2}{\pi^2 h}  [ 1-k^2
\nonumber \\
&& + (2k^2-1) C_2 - k^2 C_4 ] \biggl ] ,
\label{eq:rho-rho-a}
\ea
where we took the average over the fast oscillation period and the characteristic
density $\rho_a$ was defined in Eq.(\ref{eq:Phi-I-def}).
Because the metric function $f(r)$ goes to zero at the Schwarzschild radius,
as $f(r) \sim (r-r_s/4)^2$, the velocity $v_r \sim (r-r_s/4)^{-1}$ and the density
$\langle \rho \rangle \sim (r-r_s/4)^{-2}$ diverge at the Schwarzschild radius.
On the other hand, at large distance Eq.(\ref{eq:vr-sqrt}) gives
\be
r \gg r_{\rm sg} : \;\;\;
v_r = - \frac{F_\star m^4}{\lambda_4 \rho x^2} .
\label{eq:vr-large-r}
\ee

\subsection{Behavior at the Schwarzschild radius}
\label{sec:Schwarzschild-behavior-lambda4}

As for the case of the free scalar field studied in section~\ref{sec:free-scalar-field},
the radial velocity $v_r$ (\ref{eq:vr-sqrt}) and the density $\rho_\phi$ (\ref{eq:rho-rho-a}),
defined by the energy-momentum tensor associated with the isotropic metric,
diverge at the Schwarzschild radius because of the metric factor $1/f$.
Again, this divergence is an artifact due to the choice of coordinates, and by
going to the more appropriate Eddington metric (\ref{eq:ds2-Eddington})
we obtain finite quantities.
Thus, from Eq.(\ref{eq:vr-sqrt}) we obtain close to the Schwarzschild radius
\ba
r \to r_s/4 : && \frac{\pi \beta'}{2m} \sim - \frac{16 (1+\alpha) m r_s}{\pi (4 r - r_s)} , \\
&& \beta \sim - \frac{4 (1+\alpha) m r_s}{\pi} \ln \left( \frac{4 r - r_s}{4 r_s} \right) .
\hspace{0.5cm}
\ea
Substituting into Eq.(\ref{eq:phi-cn-def}) and using the Eddington coordinates
as in the metric (\ref{eq:ds2-Eddington}) we obtain
\be
\tilde{r} \to r_s : \;\;\; \phi = \phi_s \, {\rm cn} \left[ \frac{2 {\bf K}_s}{\pi} (1+\alpha) m
( \tilde{t}+\tilde{r} ) , k_s \right] ,
\label{eq:phi-rs-lambda4}
\ee
where the modulus $k_s$ at the Schwarzschild radius was obtained in Eq.(\ref{eq:ks-def})
and the amplitude $\phi_s$ is given by Eq.(\ref{eq:lambda4-k}) in terms of $k_s$.
As for the free scalar (\ref{eq:phi-free-rs}), the scalar field is well defined at the
Schwarzschild radius and we recover an ingoing solution with unity velocity.
However, the self-interactions remain relevant down to the Schwarzschild radius
as (\ref{eq:phi-rs-lambda4}) differs from the cosine (i.e. harmonic) expression
(\ref{eq:phi-free-rs}) of the free case. We now obtain a nonlinear radial wave,
with higher-order harmonics as given by the expansion (\ref{eq:cn-series}).

\subsection{Density profile}
\label{sec:Density-profile-lambda4}

From Eqs.(\ref{eq:rho-Eddington}) and (\ref{eq:phi-cn-def}), using
Eqs.(\ref{eq:dphi-dt-cn})-(\ref {eq:dphi-dr-cn}), the energy density
associated with the Eddington coordinates is given, at leading order in the large-$m$
limit, by
\ba
&&\hspace{-0.4cm} \tilde\rho_\phi = \frac{m^4}{\lambda_4} \frac{k^2}{1-2 k^2} \biggl \lbrace
[ 1 - k^2 + (2 k^2-1) \, {\rm cn}^2 - k^2 {\rm cn}^4 ]  \nonumber \\
&& \hspace{-0.2cm} \times \left[ 2-f + \frac{1}{f} \left( 1 - f
- \sqrt{ 1 - \frac{\pi^2 f}{(1-2 k^2) 4 {\bf K}^2 (1+\alpha)^2}} \right)^{\!\! 2} \; \right] \nonumber \\
&& \hspace{-0.2cm} \times \left( \frac{2 {\bf K} (1+\alpha)}{\pi} \right)^2
+ {\rm cn}^2 + \frac{k^2}{1-2 k^2} {\rm cn}^4 \biggl \rbrace .
\label{eq:T00-lambda4-cn}
\ea
This is the generalization of Eq.(\ref{eq:T00-free-sin}) to the case of quartic
self-interaction.
In terms of the flux $F_c$, we obtain using Eq.(\ref{eq:Fs-def}) and averaging
over the fast oscillations,
\ba
&&\hspace{-0.4cm} \langle \tilde\rho_\phi \rangle = - \frac{F_c}{F_\star r_s^2}
\frac{k^2}{1-2 k^2} \biggl \lbrace [ 1 - k^2 + (2 k^2-1) C_2 - k^2 C_4 ]  \nonumber \\
&& \hspace{-0.2cm} \times \left[ 2-f + \frac{1}{f} \left( 1 - f
- \sqrt{ 1 - \frac{\pi^2 f}{(1-2 k^2) 4 {\bf K}^2 (1+\alpha)^2}} \right)^{\!\! 2} \; \right] \nonumber \\
&& \hspace{-0.2cm} \times \left( \frac{2 {\bf K}}{\pi} \right)^2
+ \frac{1}{(1+\alpha)^2} \left( C_2 + \frac{k^2}{1-2 k^2} C_4 \right) \biggl \rbrace ,
\label{eq:T00-lambda4-F}
\ea
which generalizes Eq.(\ref{eq:T00-free-F}).
Again, this energy density remains finite at the Schwarzschild radius.
Neglecting $\alpha \ll 1$ and using $k_s \simeq 0.54$, we obtain
\be
\tilde{r} = r_s , \;\; r = \frac{r_s}{4} : \;\;\; \langle \tilde\rho_\phi \rangle \simeq
1.2 \frac{m^4}{\lambda_4} \simeq 0.9 \rho_a .
\label{eq:T00-lambda4-rs}
\ee
Contrary to the case of the free scalar, the flux $F_c$ and the density $\tilde\rho_\phi$
cannot grow arbitrarily large and take only one specific value, determined by
the self-interactions. As could be expected, the density (\ref{eq:T00-lambda4-rs})
is set by the characteristic density $\rho_a$ defined in Eq.(\ref{eq:Phi-I-def}),
which measures the strength of the self-interactions.
The unboundedness of the free case is recovered by the fact that
$\langle \tilde\rho_\phi \rangle \to \infty$ when $\lambda_4 \to 0$.

We can see that all terms in Eq.(\ref{eq:T00-lambda4-cn}) are of the same order.
This means that the terms associated with the
self-interaction potential are of the same order as those associated with the quadratic
part. Thus, close to the BH the self-interaction potential can no longer
be treated as a perturbation, which was the case on cosmological and galactic scales.
This also corresponds to the fact that the modulus $k_c$ becomes of order unity
close to the Schwarzschild radius, see (\ref{eq:ks-def}), and the Jacobi elliptic
function significantly deviates from a cosine.
Moreover, we can see that the slope of the density profile is different from the
exponent $-3/2$ obtained in the free case in Eq.(\ref{eq:rho-r-free}).
Indeed, from Eq.(\ref{eq:k-small-F}) we obtain
\be
r_s \ll r \ll r_{\rm sg} : \;\;\; k \simeq k_+ \simeq \sqrt{ \frac{2 r_s}{3 r} } .
\ee
This leads to
\be
r_s \ll r \ll r_{\rm sg} : \;\;\; \langle \tilde\rho_\phi \rangle \propto r^{-1} \;\; \mbox{and} \;\;
v_r \propto r^{-1} .
\label{eq:profile-lambda4}
\ee
As compared with the free case (\ref{eq:rho-r-free}),
the density falls off more slowly at large radii while the velocity decreases faster.

\begin{figure}
\begin{center}
\epsfxsize=8.8 cm \epsfysize=6 cm {\epsfbox{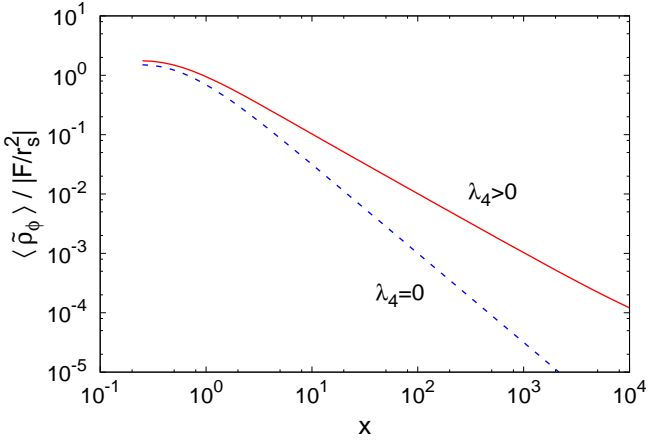}}
\end{center}
\caption{Scalar-field energy density computed in the Eddington metric, from
the Schwarzschild radius up to $10^4 r_s$, where the metric potentials are still dominated
by the central BH. We show the free case (\ref{eq:T00-free-F}) (dashed line) and the
self-interaction case (\ref{eq:T00-lambda4-F}) (solid line), for the same value $F_c$
of the flux.}
\label{fig_rho}
\end{figure}

We show in Fig.~\ref{fig_rho} the scalar field profiles of the free and interacting cases,
for the same value $F_c$ of the flux. Both densities are of the same order at the
Schwarzschild radius but we can clearly see the two different slopes for $r \gg r_s$,
with the slower falloff for the interacting case.
This corresponds in turns to a faster decay of the radial velocity.
This is not surprising, since the ``pressure'' support provided by the self-interaction
balances gravity and stabilizes the scalar-field soliton obtained at large radii,
as recalled in section~\ref {sec:large-radius-boundary}, and slows down the infall onto the
central BH at smaller radii.
On the other hand, near the Schwarzschild radius gravity cannot be resisted and the radial
velocity becomes of order unity in both cases.

\subsection{Transition radius}
\label{sec:transition}

From Eq.(\ref{eq:profile-lambda4}) we obtain the more explicit scalings
\be
r_s < r < r_{\rm sg} : \;\;\; \tilde\rho_\phi \sim \rho_a \frac{r_s}{r} , \;\;\;
v_r \sim - \frac{r_s}{r} .
\label{eq:profile-small}
\ee
This BH-dominated regime stops at the radius $r_{\rm sg}$ where the scalar field density
has decreased down to the soliton core density $\rho_s$.
This gives
\be
r_{\rm sg} = r_s \frac{\rho_a}{\rho_s} .
\label{eq:r-sg-rho}
\ee
From Eqs.(\ref{eq:Phi-BH}) and (\ref{eq:Phi-I-def}) we find at this radius
\be
r=r_{\rm sg} : \;\;\; \Phi_{\rm BH} = - \frac{\rho_s}{2\rho_a} , \;\;\;
\Phi_{\rm I} = \frac{\rho_s}{\rho_a} , \;\;\; v_r \sim - \frac{\rho_s}{\rho_a} ,
\ee
where $\Phi_{\rm BH}$ is the Newtonian potential associated with the central BH.
Normalizing the scalar-field Newtonian potential $\Phi_{\phi}$ at large radii,
beyond the soliton radius, it follows the soliton profile (\ref{eq:alpha-def}) down to $r_{\rm sg}$,
where the mass  distribution starts to deviate from the flat soliton solution.
Thus we also have
\be
r=r_{\rm sg} : \;\;\; \Phi_{\phi} = \alpha - \Phi_{\rm I} \sim - \frac{\rho_s}{\rho_a} .
\ee
Then, we can check that we indeed have $\Phi_{\phi} \sim \Phi_{\rm BH}$ at the
transition radius $r_{\rm sg}$ given by Eq.(\ref{eq:r-sg-rho}).
From Eq.(\ref{eq:vr-large-r}) we find that at larger radii, up to the soliton radius $R_s$,
we have
\be
r_{\rm sg} < r < R_s : \;\;\; \tilde\rho_\phi \sim \rho_s , \;\;\;
v_r \sim - \frac{\rho_s}{\rho_a}  \frac{r_{\rm sg}^2}{r^2} .
\label{eq:profile-large}
\ee
Of course, the spherical flux $r^2 \tilde\rho_\phi v_r$ scales as $r^0$, that is,
remains constant, in both small and large radii regimes (\ref{eq:profile-small})
and (\ref{eq:profile-large}).

For this analysis to be valid, we must check that the transition radius $r_{\rm sg}$ is smaller
than the soliton radius $R_s$. Using Eqs.(\ref{eq:r-sg-rho}) and (\ref{eq:r_a-rho_a}),
with $R_s \sim r_a$, we find that $r_{\rm sg} < R_s$ corresponds to $M < M_s$,
where $M_s \sim \rho_s r_a^3$ is the soliton mass.
The ratio $M/M_h$ of the supermassive central BH mass to the halo dark matter mass
is of order $10^{-5} - 10^{-4}$ \cite{Ferrarese:2002ct}.
On the other hand, the ratio $M_s/M_h$ of the soliton mass to the halo dark matter mass
is of order $10^{-3} - 1$ \cite{Brax:2019fzb}.
Therefore, we typically have $M \ll M_s$ and the radius $r_{\rm sg}$ that marks
the central region dominated by the BH gravitational potential is significantly
smaller than the soliton radius $R_s$.

\subsection{Scalar dark matter mass at small radii}
\label{sec:mass-small-radii}

Some scalar field dark matter models can be constrained by the measurement
of stellar dynamics at small radii, near the central supermassive BH.
For instance, an extended dark matter distribution around the BH
can affect the orbits of local stars and lead to significant precession.
This requires accurate measurements at very small radii, which start to be
available for a few cases, such as the Sgr A* BH in the  Milky Way, or the M87* BH
in the M87 galaxy. In the first case, the mass distribution is known up to the few
percent level \cite{Yu:2016nzn}; whereas for the latter one, the distribution
is constrained at the order of ten percent \cite{Akiyama:2019eap}.
This type of observations have been recently studied in this
context~\cite{Desjacques:2019zhf,Bar:2019pnz,Davies:2019wgi}.

In our case, where the scalar dark matter is supported by the self-interaction
pressure, the orders of magnitude are significantly different from the fuzzy dark matter scenario.
Let us consider the case $\rho_a \sim 1 \, {\rm eV}^4$ and $R_s \simeq 20 \, {\rm kpc}$.
For the Milky Way, with a dark matter halo mass $M_h \sim 10^{12} M_\odot$,
and a soliton mass ratio $M_s/M_h \sim 0.03 $ \cite{Brax:2019fzb},
we obtain a scalar soliton mass $M_s \simeq 3 \times 10^{10} M_\odot$.
On the other hand, the central supermassive BH has a mass
$M \simeq 4.3 \times 10^{6} M_\odot$.
This gives a Schwarzschild radius $r_s \simeq 4 \times 10^{-7} \, {\rm pc}$,
and a transition radius $r_{\rm sg} \simeq 0.1 \, {\rm pc}$.
From Eq.(\ref{eq:profile-large}), we have in the large-radius regime $r_{\rm sg} < r < R_s$
the scaling $M_\phi(<r) \propto r^3$. Therefore, we obtain at the transition radius
$M_\phi(<0.1 \, {\rm pc}) \simeq 4 \times 10^{-6} M_\odot$.
From Eq.(\ref{eq:profile-small}), we have in the small-radius regime $r_s < r < r_{\rm sg}$
the scaling $M_\phi(<r) \propto r^2$.
This gives in particular $M_\phi(<0.005 {\rm pc}) \simeq 10^{-8} M_\odot$.
The observational constraints are $M_\phi < 10^5 M_\odot$ within $0.005 \, {\rm pc}$
and $M_\phi < 10^6 M_\odot$ within $0.3 \, {\rm pc}$.
Thus, the soliton mass at small radii is far below the observational upper bounds.
On the other hand, these measurements could constrain scalar field models
such as the one studied in this paper but with very different parameters,
which would then play no role on galactic scales and only become relevant at
mpc scales.

\section{Lifetime of the scalar-field soliton}
\label{sec:lifetime}

At the typical soliton radius $r_a = R_s/\pi$, Eqs. (\ref{eq:r-sg-rho}) and
(\ref{eq:profile-large}) give for the radial velocity $v_r$ and the evolution timescale $t_c$,
\be
v_r(r_a) \sim - \frac{\rho_a}{\rho_s} \frac{r_s^2}{r_a^2}  , \;\;\;
t_c \equiv \frac{r_a}{|v_r|} \sim r_a \frac{\rho_s}{\rho_a} \frac{r_a^2}{r_s^2} .
\ee
To compare the time $t_c$ with cosmological timescales, we define the Hubble time
$t_H$ and Hubble radius $R_H$ as
\be
t_H = H^{-1} , \;\; R_H = c/H ,
\ee
and we obtain
\be
t_c \sim t_H \left( \frac{\bar\rho_c}{\rho_a} \right)^{5/2}
\frac{\rho_s}{\bar\rho_c} \left( \frac{R_H}{r_s} \right)^2 ,
\ee
where $\bar\rho_c=3H^2/(8\pi{\cal G})$ is the cosmological critical density.
This also reads at $z=0$ as
\be
t_c \sim 10^3 \, t_H \frac{\rho_s}{\bar\rho_c} \left( \frac{\rho_a}{1\,{\rm eV}^4} \right)^{-5/2}
\left( \frac{M}{10^8 M_\odot} \right)^{-2} .
\ee
For the soliton to have a radius of $20$ kpc, so that it shows a significant departure from the
CDM profiles on galactic scales, we must have $\rho_a \sim 1 \, {\rm eV}^4$
\cite{Brax:2019fzb}.
Larger characteristic densities lead to smaller soliton radii.
We typically have $\rho_s/\bar\rho_c \sim 10^5$ for the DM overdensity
in the soliton core.
Therefore, we find that $t_c \gg t_H$.
This means that the DM solitonic cores can easily survive until today, despite the
infall of their inner layers onto the central supermassive BH.

We also find that astrophysical stellar mass BHs cannot eat a significant fraction
of the galactic DM soliton.
Indeed, for $N$ BHs of unit solar mass, the typical timescale
for the soliton depletion reads
\be
t_N \sim 10^{19} \, \frac{t_H}{N} \frac{\rho_s}{\bar\rho_c}
\left( \frac{\rho_a}{1\,{\rm eV}^4} \right)^{-5/2} .
\ee
Since we typically have $N < 10^{11}$, as only a fraction of the galactic baryonic
mass can be within stellar BHs, we obtain $t_N \gg 10^8 t_H$ and
the soliton mass loss is negligible.

\section{Discussion and Conclusion}
\label{sec:conclusion}

In this work, we have analyzed steady solutions of coherent scalar fields
in galactic centers that harbor a supermassive central BH.
Neglecting the central BH, such ultralight scalar DM typically builds
a stationary coherent profile, called a soliton, with a finite radius $R_s$
and a flat core. This soliton is also embedded in
an extended halo of fluctuating density granules, with a spherically averaged
density profile that is similar to the NFW profile \cite{Navarro:1995iw}
found in numerical simulations of standard collisionless dark matter.
If $R_s$ is of the order of a few kpc, this flattened dark matter profile
can have interesting observational consequences for cosmological and galactic
studies.
In contrast with the fuzzy dark matter scenarios, with a scalar mass
$m \sim 10^{-22} \, {\rm eV}$, where the soliton is due to the
balance between gravity and the quantum pressure (associated with the wave features
of the scalar field), we focus on the case of large scalar mass,
typically $m \gg 10^{-18} \, {\rm eV}$, where gravity is instead counterbalanced
by the repulsive self-interaction associated with a quartic potential
and the quantum pressure is negligible.

In this paper, we have considered the impact of the central supermassive BH
on the profile of this soliton and its lifetime, as it gradually falls onto the BH.
As we focus on the limit of large scalar mass, we are able to perform a fully
nonrelativistic study, from the radius $R_s$ of the soliton down to the
Schwarzschild radius $r_s$. For simplicity we discard baryonic effects
but the main features of both the relativistic infall at small radii
and the soliton core at large radii should remain valid.
Baryonic matter will only increase somewhat the soliton density at intermediate
radii, where it dominates over both the central BH and scalar gravitational fields.
Then, our analysis extends from the large-radius regime $r \lesssim R_s$
dominated by the scalar dark matter self-gravity down to the small-radius regime
$r \sim r_s$ dominated by the BH gravity.
The boundary conditions at both ends determine the profile and the steady infall
onto the supermassive central BH.

First, we have studied the free massive case, associated with a quadratic scalar
potential. As the scalar field equation of motion is linear, this behaves
in a fashion similar to a collection of independent particles, with a flux
onto the central BH that is arbitrary and unbounded, proportional to the density
at large radii. As expected, at the Schwarzschild radius the scalar field takes
the form of an harmonic ingoing wave.

Then, we have extended the analysis to the self-interacting case defined by
a repulsive quartic interaction.
The limit of large scalar mass allows us to perform a fully nonlinear study,
at all orders in the coupling constant. These nonlinear dynamics generate
harmonics of all orders.
The effective pressure associated with the repulsive quartic interaction
slows down the infall onto the central BH.
In a fashion similar to the hydrodynamical case of polytropic fluids,
general relativity actually selects a unique critical value $F_c$ for the flux
of the steady infall onto the BH. This is similar to the transonic solution of the
hydrodynamical case, with a continuous switch from a low-velocity branch at large
radii, which converges to the soliton solution with a negligible radial velocity,
to a high-velocity branch at small radii, with a radial velocity that becomes
relativistic.
At the Schwarzschild radius, the scalar field takes again the form of an ingoing wave
with unit velocity, as the self-interaction pressure cannot resist the BH gravity,
but it is now a nonlinear wave that contains harmonics of all orders.
We find that in the central region, dominated by the BH gravity,
the scalar density profile and the radial velocity decay as $1/r$.
Beyond a transition radius $r_{\rm sg}$, the scalar self-gravity becomes dominant
and the scalar density follows the flat core $\rho_s$ of the soliton,
while the negligible radial velocity decays as $1/r^2$.

The critical flux $F_c$ gives a lifetime $t_c$ of the soliton that is much longer
than the age of the Universe. This implies that the soliton solutions generated
by this scalar DM scenario are not destroyed by the supermassive central BH
and are relevant.
However, because of the large soliton radius $R_s$ (as we focus on models
that could have some impact on galactic scales), the scalar dark matter mass
at small radii is very small and much below the observational upper bounds
provided by stellar dynamics close to the central supermassive BH.

By increasing the scalar mass or decreasing the quartic coupling constant,
the core density of the soliton becomes greater while its radius diminishes.
This would in turn increase the scalar density near the Black Hole and lead to stronger
effects on the stellar dynamics in this central region.
Better constraints on the mass of scalar dark matter in this regime would certainly
require to calibrate the models from large to small scales using dedicated numerical simulations,
in particular to estimate the expected mass of such small solitons.
More generally, simulations of self-interacting scalar dark matter would help
understanding the complex scalar dynamics from cosmological
scales, outside of coherent solitons, down to small subgalactic scales,
which involve soliton collisions and possible relaxation processes that are difficult
to predict in nonlinear regimes.

\section{Acknowledgements}

  This work is
supported in part by the EU Horizon 2020 research and innovation
programme under the Marie-Sklodowska grant No. 690575. This article is
based upon work related to the COST Action CA15117 (CANTATA) supported
by COST (European Cooperation in Science and Technology).
The work by JARC is partially supported by the MINECO (Spain)
project FIS2016-78859-P(AEI/FEDER, UE).

\vspace{-.3cm}

\bibliography{ref2}

%\bibliography{Biblio_scalar}

\end{document}